\documentclass[twocolumn,showpacs,preprintnumbers,amsmath,amssymb]{revtex4-1}
\usepackage{epsfig}
\usepackage{graphicx}
\usepackage{textcomp}
\usepackage{bm}
\usepackage{ulem}

\def\tr{t_{\rm r}}

\newcommand{\se}{s_{\rm ext}}
\newcommand{\0}{^{(0)}}
\newcommand{\1}{^{(1)}}

\newcommand{\Q}{{\cal Q}}

\newcommand{\pb}{{\bf p}}
\newcommand{\xb}{{\bf x}}
\newcommand{\rb}{{\bf r}}
\newcommand{\Xb}{{\bf X}}
\newcommand{\xbS}{{\bf x}_{\cal S}}

\newcommand{\Db}{\mbox{\boldmath $\Delta$}}

\newcommand{\xbo}{{\bf x}_{\rm opt}}

\newcommand{\pbo}{{\bf p}_{\rm opt}}

\def\p12{p_{12}({\bf q},t)}

\begin{document}
\title
{Control of rare events in reaction and population systems by deterministic processes and the speedup of disease extinction}

\author{M. Khasin and  M.I. Dykman}

\affiliation{Department of Physics and Astronomy, Michigan State University, East Lansing, MI 48824 USA}

\begin{abstract}
We consider control of reaction and population systems by deterministically imposed transitions between the states with different numbers of particles or individuals. Even where the imposed transitions are significantly less frequent than spontaneous transitions, they can exponentially strongly modify the rates of rare events, including switching between metastable states or population extinction. We also study optimal control of rare events, and specifically, optimal control of disease extinction for a limited vaccine supply. A comparison is made with control of rare events by modulating the rates of elementary transitions rather than imposing transitions. It is found that, unexpectedly, for the same mean control parameters, controlling the transitions rates can be more efficient.

\end{abstract}


\maketitle
\section{Introduction}

Many features of biophysical systems and population dynamics are related to the randomness of the underlying processes, such as elementary molecular reactions, or birth and death, or infection and recovery. The randomness is important where the number of involved molecules or individuals, even though it is large, is not very large, as is often the case in gene expression, for example \cite{Raser2005}. In such mesoscopic domain fluctuations are small on average. Then, if the system is in stationary conditions, for much of the time its molecular composition or the population size experience small variations about the values in dynamically stable states, i.e., the stable states that the system would have with no fluctuations. Examples are countless. Besides single-cell protein levels \cite{Rao2002,Pedraza2008}, they range from microbial populations \cite{Balaban2004} and motor proteins with different states strongly bound to the tracks \cite{Kolomeisky2007,*Sweeney2010} to, at a larger scale, insect groups and worms that can switch between the directions of motion \cite{Yates2009,*Stephens2009} and, at a still larger scale,  endemic states of infectious disease, where a part of the population is infected while the other part is not \cite{Anderson1992}, or stable states in the predator-prey models \cite{Renshaw1991}.

Even where fluctuations are small on average, occasionally there occur large fluctuations that lead to dramatic changes in the system. An example is switching between dynamically stable states. It plays an important role in biophysical systems including the ones mentioned above. Another example is extinction of a group of molecules or a population. A type of extinction that has attracted much attention in the literature is spontaneous disease extinction in an isolated population, where as a result of a fluctuation the number of infected becomes equal to zero and the epidemics ceases \cite{Weiss1971,Leigh1981,Herwaarden1995,Andersson2000,Nasell2001,Elgart2004,Doering2005,Assaf2006,Kessler2007,Dykman2008,Kamenev2008,Shaw2010,Adams2010}.

In this paper we will be interested in the control of the rates of rare events that result from large fluctuations, including interstate switching and extinction. We will develop a general approach to control by deterministically imposed elementary transitions, in which the numbers of molecules or individuals change. The change is small and does not cause switching or extinction on its own. We show that, nevertheless, even weak deterministic control can strongly increase the rates of rare events. The approach will then be applied to control of disease extinction. Here control is often performed by vaccination. Sometimes there is not enough vaccine to eradicate the disease. We show how to optimize the control given a restriction on the amount of vaccine. The results refer also to a more general situation of a limited time-average speed at which the transitions are imposed.

The problem of optimal control of large rare fluctuations has been discussed in qualitatively different terms for two major models of fluctuating systems, continuous and discrete. Fluctuations in continuous systems are often thought of as resulting from an external noise, whereas control of such systems is performed by an applied regular field. This field can be optimized, for a given constraint, so that it will most efficiently change the rate of noise-induced rare events \cite{Smelyanskiy1997,Vugmeister1997}.

Populations or reacting species, on the other hand, are intrinsically discrete, with integer numbers of reacting molecules or individuals. The dynamics is often described as resulting from elementary transitions in which these numbers change. The transitions happen at random and are characterized by rates. These rates can be controlled, but the transitions still remain random. We call this elementary transition rates (ETR) control. Optimal control of this type was discussed in Ref.~\onlinecite{Khasin2010a}. Because of the randomness of the elementary transitions, the constraint on the control field is in some sense probabilistic, for example, it can be imposed on the average amount of vaccine used in a given time period.

We will consider here a different type of control, which is performed via deterministic processes (DP) where the numbers of reacting objects (molecules or individuals) change in a well-defined way at well-defined instants of time.  We call it DP-control. We will show how to incorporate deterministic transitions into the analysis of large rare fluctuations. In particular, we will develop an eikonal approximation, which applies where the total numbers of reacting objects are large. It maps the problem of the probabilities of rare events in a reaction system onto Hamiltonian dynamics of an auxiliary system. This formulation will be compared with the corresponding formulation for controlled reaction rates and for controlled continuous systems.

The approach will then be applied to the problem of disease extinction with a limited amount of vaccine. We show that periodic vaccination implemented in a deterministic fashion can strongly enhance disease extinction. The enhancement can be resonant, for the appropriate vaccination period.

\section{Master equation in the presence of deterministic processes}

The state of the reaction or population system is determined by vector $\Xb=(X_1,X_2,...)$ with integer-valued components that are equal to the numbers of molecules of different sorts or individuals in different population groups. The size of the system $N$, which gives the typical value of $|\Xb|$, is the large parameter of the theory, $N\gg 1$. Our analysis refers to spatially uniform systems, which is of interest for many applications in biophysics, stirred chemical systems, and also for population dynamics in moderately large globally connected groups.

\begin{figure}[ht]
\includegraphics[width=8.5truecm]{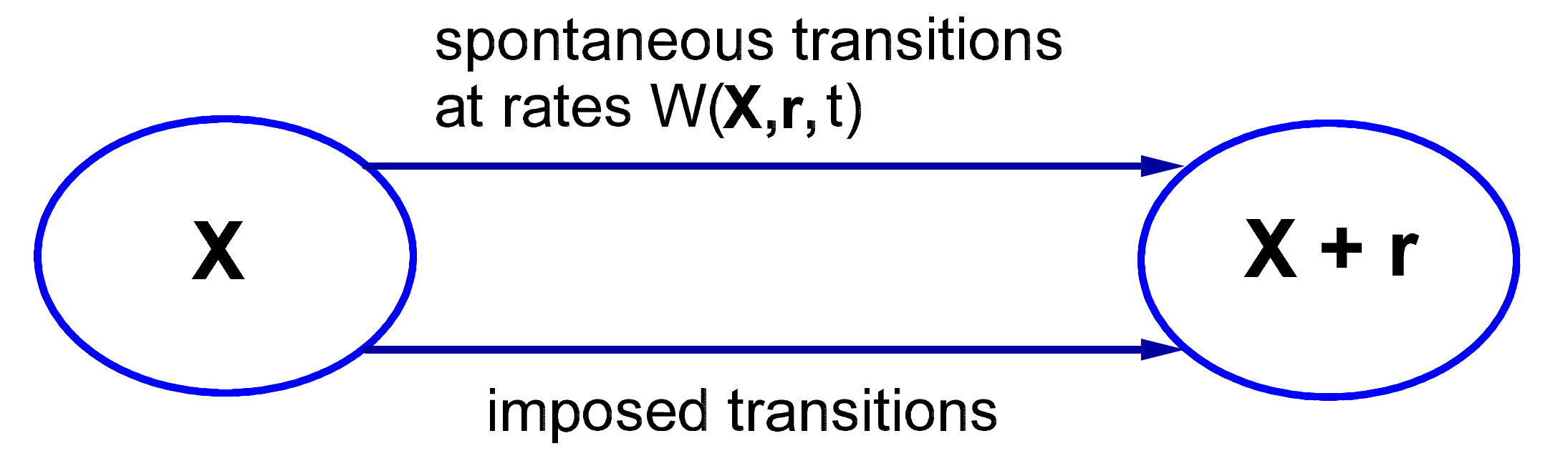}
\caption{Two types of transitions with a change of the number of molecules or individuals $\Xb\to \Xb+{\bf r}$. Spontaneous transitions happen at random and are characterized by rates $W(\Xb,\rb,t)$. In addition, transitions may be imposed externally in a deterministic way. We assume that the duration of a transition is small compared to the interval between successive transitions, so that the system is described by vector $\Xb$ with integer-valued components.}
\label{fig:formulation}
\end{figure}

We will be interested in the probability $P(\Xb,t)$ for the system to be in state $\bf X$. The distribution $P$ changes as a result of elementary transitions in which $\Xb$ is changed. Such transitions can come from spontaneous chemical or biochemical reactions, or from infection and recovery of individuals, for example. They can be imposed also in a deterministic fashion, see Fig.~\ref{fig:formulation}. We will assume that, for a given molecule or an individual, the interval between successive transitions, whether spontaneous or deterministic, is much longer than the duration of a transition. Then elementary transitions can be considered instantaneous, which is consistent with the notion of the components of $\Xb$ being integer-valued.

In the absence of deterministic transitions the evolution of $P(\Xb,t)$ is described by the standard Markov master equation
\begin{eqnarray}
\label{eq:master}
\partial_t P(\Xb,t)&=&\hat{W}  P(\Xb,t),
\end{eqnarray}
where
\begin{eqnarray}
\label{eq:rate_operator}
\hat{W}  P(\Xb,t)&=& \sum_{\rb}\left[W(\Xb-\rb, \rb,t)P(\Xb - \rb,t)\right. \nonumber \\
&-&\left. W(\Xb, \rb,t)P(\Xb,t)\right].
\end{eqnarray}
Here, $W(\Xb,\rb,t)$ is the rate of an elementary transition $\Xb\to \Xb+\rb$ in which the population and/or number of molecules changes by $\rb=(r_1,r_2,\ldots)$.  The rates $W$ are proportional to the system size, $W\propto N$. The change of $\Xb$ in an elementary transition is independent of $N$ and comparatively small, $|\rb|\ll N$, typically $|\rb|\sim 1$. We will assume that the rates $W$ are either independent of time or depend on time periodically.

We now discuss the effect of a control field that deterministically imposes elementary transitions. This field is a series of pulses applied at instants $t_1<t_2...<t_n<\ldots$. In each pulse the state vector changes by $\Db$, that is, $\Xb\to \Xb+ \Db$. We will be interested primarily in the case where $|\Db|\sim 1$. An example is vaccination in population dynamics. One can think that vaccination shots are applied at instants $t_n$. In a simplified model where the delay between vaccination and acquiring immunity is disregarded, as a result of a shot the number of susceptible individuals decreases by one, whereas the number of vaccinated increases by one. The corresponding components of vector $\Db$ are then -1 and 1, respectively, whereas all other components are equal to zero.

The analysis should be performed differently if a change that occurs in a pulse is macroscopic, that is if $|\Db|$ is of order $N$. This may be of interest, for example, if control is performed by injecting or extracting a macroscopic amount of reacting molecules. The corresponding extension is provided at the end of Sec.~\ref{sec:eikonal}.

Disregarding the pulse duration, one can describe the effect of the deterministically imposed pulses as a shift of the probability distribution,
\begin{eqnarray}
\label{eq:jump}
P(\Xb + \Db,t_n+0)&=&P(\Xb,t_n-0).
\end{eqnarray}
The formulation can be immediately generalized also to the case where the population change in a pulse $\Db$ depends on the instant $t_n$. We will be primarily interested in the case where the inter-pulse intervals are larger than the typical reciprocal reaction rate.

In the presence of a pulsed deterministically acting field, master equation (\ref{eq:master}) applies in between the pulses. Equations (\ref{eq:master}) - (\ref{eq:jump}) fully describe the evolution of the distribution $P(\Xb,t)$. However, because of the discontinuity of $P$ at instants $t_n$, they are inconvenient for the analysis.

\section{The eikonal approximation}
\label{sec:eikonal}

For a large system and $|\Db|\sim 1$, the typical inter-pulse interval $t_n-t_{n-1}$ is small, $\propto 1/N$, since the control field is applied to different individuals. If this were not the case, the overall effect of the field would be negligibly small. On the other hand, of interest is the evolution of the distribution $P$ on much longer times, of the order of the relaxation time of the system $\tr$. The time $\tr$ characterizes how the system approaches the stable state in the absence of fluctuations and is independent of the system size.

Describing evolution of $P$ on times $\sim\tr$ requires coarsening over time intervals $\propto 1/N$, and in particular over the inter-pulse interval. This can be done by introducing the characteristic pulse speed $\xi(t)$,
\begin{eqnarray}
\label{eq:speed_xi}
\xi(t_n)=\frac{1}{N (t_n-t_{n-1})}.
\end{eqnarray}
The continuous function $\xi(t)$ is obtained by analytic continuation from the discrete set $\xi(t_n)$; it is assumed that $\xi(t)$ is smooth, $\dot\xi(t)(t_n-t_{n-1})\ll \xi(t)$ for $|t-t_n|\ll \tr$. Note that we are calling $\xi(t)$ {\it speed} in contrast to {\it rate} to emphasize its deterministic nature and to distinguish it from the rates used in other equations.

Function $\xi(t)$ determines the number of pulses $N\xi(t)\tau$ that happen in a time interval $\tau$ such that $t_n-t_{n-1}\ll \tau\ll \tr$. We will assume that $\xi(t)$ is either constant or periodic in time, with period $T \gg t_n-t_{n-1}$. We further assume, for simplicity, that this period coincides with the period of the rates $W(\Xb,\rb,t)$ if these rates are time-dependent. Then, in the neglect of fluctuations, the system can have periodic steady states; the results can be immediately extended to the case where the periods of $W$ and $\xi$ are commensurate.

Introducing $\xi(t)$ does not allow one to directly coarsen the master equation for the probability distribution $P(\Xb,t)$ over time. Indeed, on the tail of the distribution, which is of interest for the problem of large rare fluctuations, function $P$ is steep, $|P(\Xb+\Db,t)-P(\Xb,t)|\sim P(\Xb,t)$, which means that it significantly changes from pulse to pulse and thus $P(\Xb,t+\tau)-P(\Xb,t)$ is not given by $\tau\partial_t P$ even for $\tau\ll \tr$.

\subsection{The Hamiltonian formalism}

A differential equation that describes important features of the evolution of the distribution in coarsened time can be obtained by seeking $P(\Xb,t)$ in the eikonal form \cite{Kubo1973,Gang1987,Dykman1994d},
\begin{eqnarray}
\label{eq:eikonal}
P(\Xb,t)=\exp[-N s(\xb,t)], \qquad \xb\equiv  \Xb/N,
\end{eqnarray}
where $\xb$ is the scaled state vector. In the limit $N \gg 1$ this vector is quasi-continuous. We assume that function $s$ is a smooth function of $\xb$ and $t$. This assumption will be checked {\it a posteriori}.

To obtain an equation for $s$ between the pulses one can write the rate operator in Eq.~(\ref{eq:rate_operator}) as
\begin{eqnarray*}
\hat WP= \sum\nolimits_{\rb}[\exp(-\rb\partial_{\Xb})-1]W(\Xb,\rb,t)P(\Xb,t).
\end{eqnarray*}
Functions $W(\Xb,\rb,t)$ usually smoothly (polynomially) depend on $\Xb$. In contrast, function $P$ can be exponentially steep, as seen from Eq.~(\ref{eq:eikonal}). Respectively, to the leading order in $1/N$ one should only differentiate $P$ in the above expression. Then from Eqs.~(\ref{eq:master}) and (\ref{eq:jump}) one obtains
\begin{eqnarray}
\label{eq:action_w_boundaries}
&&s(\xb+\frac{1}{N}\Db,t_n+0)=s\left(\xb,t_n-0\right),  \nonumber \\
&&\partial_t s=-H_w(\xb,\partial_{\xb}s,t), \qquad \ t_n<t<t_{n+1}
\end{eqnarray}
where $n=1,2,\ldots$,
\begin{eqnarray}
\label{eq:hamiltonian}
 H_w(\xb,\pb,t)&=&\sum\nolimits_{\bf r}w(\xb,\rb,t)\left[\exp(\pb\rb)-1\right], \nonumber \\
  w(\xb,\rb,t)&\equiv& \frac{1}{N} W(N\xb,\rb,t).
\end{eqnarray}
The terms disregarded in the equation for $s$ between the pulses give corrections $\propto 1/N$ \cite{Escudero2009}.

The change of $s$ resulting from a single pulse is small, $\propto 1/N$ [at the same time, from Eq.~(\ref{eq:eikonal}), $P$ can change significantly]. This allows us to obtain a single differential equation for $s$ on a coarsened time scale. Indeed, choosing time $\tau$ so that $t_n-t_{n-1}\ll \tau \ll \tr$, in which case many pulses occur during the time $\tau$, but the overall change of $s$ accumulated between the pulses is small, we obtain
\begin{eqnarray}
\label{eq:action_intermediate}
s(\xb,t+\tau)-s(\xb,t)&\approx &-\tau H_w(\xb,\partial_{\xb}s(\xb,t),t)\nonumber \\
&-& \tau \xi(t) \Db\cdot\partial_{\xb} s,
\end{eqnarray}
where $\xi(t)$ is given by Eq.~(\ref{eq:speed_xi}). We can now formally go to the limit $\tau \rightarrow 0$, which gives
\begin{eqnarray}
\label{eq:HJ_regular_force}
\partial_t s(\xb,t)&=&- H_w(\xb,\partial_{\xb}s(\xb,t),t)- \xi(t) \Db\cdot\partial_{\xb} s.
\end{eqnarray}

Equation (\ref{eq:HJ_regular_force}) has the form of a Hamilton-Jacobi equation, with $s$ being the action of an auxiliary dynamical system with coordinate $\xb$, momentum $\pb=\partial_{\xb}s$, and Hamiltonian
\begin{eqnarray}
\label{eq:hamiltonian total}
 H(\xb,\pb,t)=H_w(\xb,\pb,t) + \xi(t) \pb \Db.
\end{eqnarray}
Hamiltonian $H$ contains both the well-known part $H_w$ that comes from random elementary transitions \cite{Wentzell1976,Gang1987,Dykman1994d} and a part that comes from the deterministic processes. The latter is described by the last term in $H$ and is characterized by two natural parameters: the speed of transitions per individual $\xi(t)$ and the change in population at a transition $\Db$. The structure of this term is very different from that of the term $H_w$: it depends linearly  on the momentum of the auxiliary system $\pb$, whereas $H_w$ depends on $\pb$ exponentially, that is, much stronger. Since functions $w,\xi$ are either independent of time or periodic with the same period, Hamiltonian $H$ as a whole is either independent of time or depends on time periodically.

The reduction of the problem of the distribution tail to the Hamilton-Jacobi equation (\ref{eq:HJ_regular_force}) makes it possible to study the distribution in the presence of deterministic control using standard tools of the rare-event theory \cite{Wentzell1976,Gang1987,Dykman1994d}.
The assumed smoothness of $s$ as function of $\xb$ immediately follows from Eq.~(\ref{eq:HJ_regular_force}) for smooth initial conditions.

\subsection{Comparison with the previously studied models}

It is instructive to compare Eqs.~(\ref{eq:HJ_regular_force}) and (\ref{eq:hamiltonian total}) with the description of large fluctuations in noise-driven continuous systems in the presence of a control field. There, for white Gaussian noise, the probability density can be also sought in the eikonal form of Eq.~(\ref{eq:eikonal}), with $N$ replaced by the reciprocal noise intensity \cite{Freidlin_book}. The Hamiltonian has a structure similar to Eqs.~(\ref{eq:hamiltonian}) and (\ref{eq:hamiltonian total}), with the difference being that $\exp(\pb\rb)$ in $H_w$ should be expanded to second order in $\pb$. The form of the control-field induced term is the same as in Eq.~(\ref{eq:hamiltonian total}), if the field corresponds to an additive force driving the continuous system, with $-\xi(t)\Db$ playing the role of this force, cf. \cite{Smelyanskiy1997,Vugmeister1997}.

We should also compare the present formulation with that for the elementary transition rates control \cite{Khasin2010a}. The type of the ETR control most close to the considered here DP control is where elementary transitions $\Xb\to \Xb + \Db$ rather than being directly imposed occur at random at a controlled rate $N\xi_{\rm pr}(t)$ which is independent of $\Xb$ \cite{Dykman2008}. As we will see, in the mean-field approximation where fluctuations are disregarded, the evolution of the system in the presence of such control is described by the same equation as for the DP control.

To account for the ETR control, one should incorporate the term describing the controlled transitions in the operator $\hat W$ in Eq.~(\ref{eq:master}), $W(\Xb,\rb,t)\to W(\Xb,\rb,t)+ N\xi_{\rm pr}(t)\delta_{\rb,\Db}$. In the eikonal approximation, to the leading order in $1/N$ one then obtains the Hamilton-Jacobi equation for $s(\xb,t)\approx -N^{-1}\ln P(N\xb,t)$, with Hamiltonian
\begin{equation}
\label{eq:hamiltonian_transitions}
H_{\rm pr}(\xb,\pb,t)=H_w(\xb,\pb,t)+ \xi_{\rm pr}(t)[\exp(\pb \Db)-1].
\end{equation}
The major distinction from the DP control, Eq.~(\ref{eq:hamiltonian total}), is in the different dependence of the control term on momentum $\pb$. The origin of this difference and its consequences for the control will be discussed in Sec.~\ref{sec:mean_field}. Formally, the DP-term in the Hamiltonian $H$, Eq.~(\ref{eq:hamiltonian total}), looks like a small-$\pb\Db$ expansion of the term $\propto \xi_{\rm pr}$ in $H_{\rm pr}$.

\subsection{Simultaneously imposed multiple transitions}

The analysis above should be modified if the control pulses are applied rarely but cause macroscopic changes in the system, so that a {\it macroscopic portion} of the molecules or population is changed in a single pulse. Formally, this means that $|\Db|\gg 1$ and the ratio $|\Db|/N$ is independent of $N$ in the limit of large $N$. As mentioned above, an example is provided by injection or extraction of a macroscopic number of molecules of a certain sort into the chemical reactor or the biological cell, or a group of individuals into the population. If the pulses are short, one can describe their effect by Eq.~(\ref{eq:jump}). We will assume that $|\Db|/N$ is sufficiently small, so that the system is weakly changed by a single pulse.

For the system to have a steady periodic state, the pulses should be applied periodically, with $t_n-t_{n-1}=t_{n+1}-t_n\sim \tr$. The tail of the distribution can still be sought in the eikonal form, Eq.~(\ref{eq:eikonal}). However, now the equation for $s(\xb,t)$ is of the form
\begin{eqnarray}
\label{eq:multiple_transitions_Hamiltonian}
\partial_ts&=&-H_w(\xb,\partial_{\xb}s,t)\nonumber\\
 &&+ \sum_n\delta(t-t_n)\left[\exp\left(N^{-1}\Db\cdot\partial_{\xb}\right)-1\right]s.
\end{eqnarray}
It differs from the Hamilton-Jacobi equation, and the general analysis of this equation is beyond the scope of this paper. However, we are interested here in a comparatively weak deterministic control, where $|\Db|/N$ is small. Then it is sufficient to keep the lowest-order term in $|\Db|/N$ in Eq.~(\ref{eq:multiple_transitions_Hamiltonian}), and this equation becomes of the same form as Eq.~(\ref{eq:HJ_regular_force}), with $\xi(t)$ in Eq.~(\ref{eq:HJ_regular_force}) of the form of a periodically repeated $\delta$-function.

\section{Mean field dynamics and optimal fluctuations}
\label{sec:mean_field}

The action $s(\xb,t)$, and thus the distribution $P$ to the leading order in $1/N$ can be found from the Hamiltonian equations of motion that follow from Eq.~(\ref{eq:hamiltonian total}),
\begin{eqnarray}
\label{eq:eom_Hamiltonian}
&&\dot\xb=\sum\nolimits_{\rb}\rb w(\xb,\rb,t)e^{\pb\rb} + \xi(t)\Db, \nonumber\\
&&\dot \pb=-\sum\nolimits_{\rb}\partial_{\xb} w(\xb,\rb,t)\left(e^{\pb\rb}-1\right).
\end{eqnarray}
These equations have an important solution
\begin{eqnarray}
\label{eq:mean_field}
\dot{\bar\xb} = \sum\nolimits_{\rb}\rb w(\bar\xb,\rb,t)+ \xi(t)\Db, \quad \pb={\bf 0}.
\end{eqnarray}
It describes the mean-field dynamics of the population, i.e., the dynamics in the neglect of fluctuations. Equation (\ref{eq:mean_field}) can be obtained also directly from Eqs.~(\ref{eq:master}) and (\ref{eq:jump}) by multiplying them by $\Xb$, summing up over $\Xb$ while disregarding the width of the distribution $P(\Xb,t)$, and coarsening over time. In the case of the ETR control described by Hamiltonian (\ref{eq:hamiltonian_transitions}), the equation of motion for $\bar \xb$ has the same form as Eq.~(\ref{eq:mean_field}), except that $\xi(t)$ is replaced by $\xi_{\rm pr}(t)$.

We assume that Eq.~(\ref{eq:mean_field}) has a stable solution $\xb_A$. We will consider the simple and fairly common case where $\xb_A$ is stationary, for time-independent rates $w(\xb,\rb,t)$ and constant $\xi(t)$, whereas for periodically varying in time rates and/or periodic $\xi(t)$ (with the same period), $\xb_A$ is periodic with the same period as the modulation. The time $\tr$ gives the relaxation time of $\bar \xb$ and can be found from Eq.~(\ref{eq:mean_field}) linearized about $\xb_A$.

We also assume that Eq.~(\ref{eq:mean_field}) has an unstable saddle-type stationary or periodic state $\xbS$ on the boundary of the basin of attraction to $\xb_A$. In the problem of extinction, at $\xbS$ one of the types of molecules or population groups becomes extinct. Respectively, one of the components of vector $\xb$, which we call $x_E$, becomes equal to zero. The state $\xbS$ is stable with respect to all components $x_{i\neq E}$. Note that, by construction, $x_i\geq 0$, the system cannot go beyond the extinction state to negative $x_E$. In addition we assume that, once the group has gone extinct, it does not re-emerge, that is, once the hyperplane $x_E=0$ has been reached, the system will stay there.

In the problem of switching between coexisting stable states, on the other hand, $\xbS$ is a conventional saddle state. In this case, generally $(\xbS)_i>0$ for all $i$ and the system can be on any side of $\xbS$ along any component $x_i$. From the opposite sides of $\xbS$ in the unstable direction the system will go to different stable states, in the neglect of fluctuations.

We assume that there is only one state $\xbS$ on the boundary of the basin of attraction to $\xb_A$. The system can either switch or one group of molecules or population can go extinct. The control field can shift the equilibrium position $\xbS$. In the extinction problem, the unstable state should be in the hyperplane $x_E=0$. The control field cannot lead to the re-birth of the group that has gone extinct. Therefore in the considered model of imposed transitions $\Delta_E=0$ for the extinction problem.

\subsection{Logarithmic susceptibility}

Equations (\ref{eq:eom_Hamiltonian}) can be used to find the rates of fluctuation-induced switching and extinction in biophysical and population systems. The problems of calculating these rates have much in common, but are not identical. The difference between them comes from the fact that in switching the system has to cross the boundary of the basin of attraction of the initially occupied stable state $\xb_A$ and go to the other state, whereas in extinction it suffices to reach the hyperplane $x_E=0$, and then the system will stay in this hyperplane. This difference shows in Eqs.~(\ref{eq:switching_bounadry}) and (\ref{eq:boundary}).

To logarithmic accuracy, the extinction and switching rates $W_e$ are determined by the probability distribution $P(\Xb,t)$ for $\xb\to \xbS$ and $W_e^{-1}\gg t\gg \tr$, given that initially the system was in the vicinity of state $\xb_A$ \cite{Dykman1994d,Dykman2008}, with $\xbS$ being the extinction and saddle states, respetively. Taking into account that, in this time range, the distribution is maximal at $\xb_A$, we have \cite{Dykman1994d,Elgart2004,Dykman2008,Kamenev2008}
\begin{eqnarray}
\label{eq:define_Q}
W_e\propto \exp(-{\cal Q}), \qquad {\cal Q}=N\se, \\
\se=\int_{-\infty}^{\infty}dt\left[\pb\dot\xb-H(\xb,\pb,t)\right].\nonumber
\end{eqnarray}
The quantity $\se$ is given by the action $s(\xb)$ for $\xb\to \xbS$ counted off from the value of $s$ at the stable state $\xb_A$. It can be calculated by minimizing the functional $\se$ with respect to $\bigl(\xb(t),\pb(t)\bigr)$, which leads to finding the Hamiltonian trajectory (\ref{eq:eom_Hamiltonian}) that starts for $t\to -\infty$ at state $\xb_A$ and approaches state $\xbS$ for $t\to \infty$. This optimal trajectory, $(\xbo(t),\pbo(t))$, gives the most probable path followed by the system in spontaneous extinction or switching.

The boundary conditions for the switching and extinction problems were discussed in \cite{Dykman1994d,Dykman2008,Khasin2010a}. One can extend the analysis to show that, in the presence of the DP control that we consider, these boundary conditions do not change. Specifically, in the switching problem
\begin{equation}
\label{eq:switching_bounadry}
\pb\to 0, \qquad t\to \pm\infty,
\end{equation}
whereas in the extinction problem
\begin{equation}
\label{eq:boundary}
\pb\to 0, \; t\to -\infty; \qquad p_{i\neq E}\to 0,\; t\to \infty,
\end{equation}
while the component $p_E$ remains nonzero for $t\to\infty$. In the both problems, the Hamiltonian $H\to 0$ for $t\to \pm\infty$.

Of utmost interest for us is the effect of weak periodic control,
\begin{equation}
\label{eq:weak_control_condition}
\xi(t+T)=\xi(t),\qquad \xi(t)\tr\ll 1,
\end{equation}
where $T$ is the modulation period. In this case $\se$ has a simple form $\se= \se\0+ \se\1$, where
\begin{eqnarray}
\label{eq:action_w_LS}
&&\se\1[\xi(t)]=\min_{t_0}\int\nolimits_{-\infty}^{\infty}dt \chi(t-t_0)\xi(t),\\
&&\chi(t)=-\pbo\0(t)\Db.\nonumber
\end{eqnarray}
Here, $\se\0$ is the action in the absence of the control field, and $\bigl(\xbo\0(t),\pbo\0(t)\bigr)$ is the most probable (optimal) extinction or switching path for $\xi=0$; $\se\1$ gives the field-induced correction to the action. We are interested in the case where $\se\1 < 0$; only in this case the control increases $W_e$.

Equation (\ref{eq:action_w_LS}) is written for the case where the elementary transition rates $W(\Xb,r)$ are independent of time. In this case the fluctuation leading to extinction or switching can happen at any time with the same probability; respectively, the optimal path $\bigl(\xbo\0(t),\pbo\0(t)\bigr)$ can be centered (for example, $|\dot{\xb}_{\rm opt}\0(t)|$ can reach maximum) at an arbitrary time $t_0$. A time-dependent control lifts this time degeneracy. It synchronizes the optimal path so as to most strongly decrease action $\se$. This synchronization is formally described in Eq.~(\ref{eq:action_w_LS}) by minimization with respect to $t_0$.

Even where the correction to the action is small, $|\se\1| \ll \se\0$, the change of the switching or extinction rate, which is given by the factor $\exp(-N\se\1)$, can be very large. It is this large change that makes the control exponentially efficient. Function $\chi(t)$ is called the logarithmic susceptibility \cite{Smelyanskiy1997c}. It describes the linear response of the logarithm of the probability $P(\Xb)$ to the control field.

Equation (\ref{eq:action_w_LS}) describes also the effect of simultaneously imposed multiple transitions, where $|\Db|\gg 1$, provided $|\pbo\0(t)\Db|/N \ll 1$. Function $\xi(t)$ in Eq.~(\ref{eq:action_w_LS}) should be replaced with $-N^{-1}\sum_n\delta(t-t_n)$ in this case.

If the elementary transition rates $W(\Xb,\rb,t)$ are periodic in time, the optimal trajectories $\left(\xbo\0(t),\pbo\0(t)\right)$ are periodically repeated in time, but their phase is fixed by the time dependence of $W(\Xb,\rb,t)$. If $\xi(t)$ has the same period, the correction to the action is
%
\[ \se\1[\xi(t)]=\int\nolimits_{-\infty}^{\infty}dt \chi(t)\xi(t) \]
%
%
with $\chi(t)$ of the same form as in Eq.~(\ref{eq:action_w_LS}). In this case, however, the control field does not synchronize transitions and its effect depends on its phase with respect to the optimal trajectories of extinction or switching.

\subsection{Comparison with the control of elementary transition rates}

The analysis of the weak ETR control was done earlier\cite{Dykman2008,Khasin2010a}. It is based on the eikonal formulation with the Hamiltonian (\ref{eq:hamiltonian_transitions}). The control-induced change of the extinction or switching rate $W_e$ is given by Eqs.~(\ref{eq:define_Q}) and (\ref{eq:action_w_LS}) in which $\xi(t)$ is replaced with $\xi_{\rm pr}(t)$ and the logarithmic susceptibility $\chi(t)$ is replaced with
\begin{equation}
\label{eq:chi_probabilistic}
\chi_{\rm pr}(t)=1-\exp[\pbo\0(t)\Db].
\end{equation}
By comparing Eqs.~(\ref{eq:action_w_LS}) and (\ref{eq:chi_probabilistic}) one can see that $-\chi_{\rm pr}\geq -\chi$, for the same $\Db$. This shows that the sensitivity of the extinction rate to the ETR control is higher than to the DP control.

The stronger effect of the ETR control for the same rate $\xi_{\rm pr}(t)$ as the speed $\xi(t)$ and for the same change of the number of molecules or individuals in a transition $\Db$ is somewhat unexpected. It can be understood qualitatively in the following way. In both cases the control field does not cause extinction on its own. It cooperates with the fluctuation that leads to extinction. The stronger the field the stronger is its effect, but also the effect increases with the increasing momentum on the optimal extinction trajectory $\pb =\partial_{\xb}s$. This momentum gives the steepness of the probability distribution, as seen from Eq.~(\ref{eq:eikonal}). The steeper the distribution the stronger is the effect of changing it by the control field.

For the ETR control, the control-induced transitions have the Poisson distribution. There is a probability that within a given time interval $\tau$ this number will be higher than the average number $N\xi_{\rm pr}(t)\tau$. Where the distribution in the absence of control is steep, that is, $|\pb|$ is large, the major effect on $W_e$ will come from realizations of the control with such more frequent transitions.

In other words, for the ETR control one can think of extinction or switching as resulting from two types of fluctuations: fluctuations in the absence of the control and fluctuations of the control field itself. The extinction rate $W_e$ is determined by the optimal, most probable fluctuations of the both types. It is this double optimization that makes $W_e$ more sensitive to the ETR control.

\section{Optimal control by deterministic transitions: vaccination for a limited vaccine supply}

We now consider the problem of optimal control of extinction and switching rates by deterministically imposed transitions. To be specific, we consider disease extinction, with the control performed by vaccination. The results are not limited to this model, the only condition that we use is that the speed $\xi(t)\geq 0$. This condition follows from Eq.~(\ref{eq:speed_xi}). We also assume that $\xi(t)$ is periodic. The optimal form of $\xi(t)$, i.e., the optimal control protocol, depends on the imposed constraints. The constraint that we consider is fairly general, and again, is not limited to vaccination only. Essentially, we consider a constraint on the number of imposed transitions per modulation period.

As mentioned earlier, a simple model of vaccination is where the vaccine is applied to susceptible individuals, and in an elementary transition a susceptible becomes vaccinated and thus immediately immune to the infection. A natural constraint on the speed of vaccination is that the total amount of vaccine per period $N\Xi$ is fixed,
\begin{eqnarray}
\label{eq:mean_vaccine_rate}
T^{-1} \int\nolimits_0^Tdt\,  \xi(t) =\Xi,
\end{eqnarray}
where $T$ is the vaccination period, $\xi(t+T)=\xi(t)$.

The optimal vaccination speed $\xi(t)$ should minimize the disease extinction barrier $\Q$ subject to constraint (\ref{eq:mean_vaccine_rate}). From Eq.~(\ref{eq:define_Q}), optimal $\xi(t)$ is given by the solution of the variational problem of minimizing the functional
\begin{eqnarray}
\label{eq:var_func}
\tilde s_{\rm ext}\left[\xi(t)\right]=\se\left[\xi(t)\right]+\lambda T^{-1}\int\nolimits_0^T \left[\xi(t)-\Xi\right]dt,
\end{eqnarray}
where $\lambda$ is the Lagrange multiplier and $\se[\xi]$ is given by Eq.~(\ref{eq:define_Q}).

For comparatively weak vaccination, where $\xi$-dependence of $\se$ is of the form of Eq.~(\ref{eq:action_w_LS}), minimization with respect to $\xi(t)$
can be done in the same way as for probabilistic vaccination where the vaccination is a Poisson process characterized by rate \cite{Khasin2010a}; such vaccination is an example of the ETR control. It follows from the form of the constraint (\ref{eq:mean_vaccine_rate}) and the condition $\xi(t)\geq 0$ that the optimal vaccination speed as a function of time is independent of the logarithmic susceptibility $\chi(t)$. If the system is stationary in the absence of periodic vaccination, from Eq.~(\ref{eq:action_w_LS})
\begin{eqnarray}
\label{eq:inequal_se}
&&\se\1\geq \min_t\chi_T(t)\int_0^Tdt \xi(t)=\Xi T\min_{0\leq t<T}\chi_T(t),\nonumber\\
&&\chi_T(t)=\sum_{n=-\infty}^{\infty}\chi(t+nT).
\end{eqnarray}
The minimum of $\se\1$ is reached, and the inequality (\ref{eq:inequal_se}) becomes an equality, for $\xi(t)$ of the form of periodically repeated $\delta$-pulses,
\begin{eqnarray}
\label{eq:action_minimized_explicit}
\xi(t)=\Xi \,T\sum_n \delta (t-t_{\min}+nT).
\end{eqnarray}
The instant $t_{\min}$ determines where vaccination pulses are applied. It is arbitrary for a stationary system. In fact, the disease extinction events are adjusted to the vaccination pulses. This adjustment is described by the minimization over $t$ in Eq.~(\ref{eq:inequal_se}); the periodic function $\chi_T(t)$ is minimal at $t_{\min}$.

The optimal shape of $\xi(t)$ is easy to understand: the vaccine is most efficient if it acts where $|\chi_T(t)|$ is maximal, and also where $\chi_T(t)<0$ to assure that $\se\1<0$. We note that strong periodic vaccination has been investigated in the framework of  deterministic epidemic models, where all fluctuations are disregarded, and it was found that pulsed vaccination is advantageous compared to vaccination at a constant rate \cite{Shulgin1998}.

For periodically modulated systems, where $W(\Xb,\rb,t)$ are periodic functions of time, optimal $\xi(t)$ still has the form of pulses, Eq.~(\ref{eq:action_minimized_explicit}), but now spontaneous extinction events are synchronized without vaccination, and this is the vaccination that must adjust in order to increase the rate of disease extinction. Equation (\ref{eq:inequal_se}) for $\se\1$ applies only if $t_{\min}$ is chosen so that $\chi_T(t)$ is minimal at $t_{\min}$; a wrong choice of $t_{\min}$ will be less efficient and can even slow down spontaneous disease extinction instead of speeding it up.

It is instructive to compare the results with  the probabilistic ETR-type vaccination, where vaccination is applied at random with rate $\xi_{\rm pr}(t)$. The differences are in the form of the logarithmic susceptibility, cf. Eqs.~(\ref{eq:action_w_LS}) and (\ref{eq:chi_probabilistic}), and in the meaning of the parameter $\Xi$. In the deterministic scenario, $NT\Xi$ gives the actual number of individuals vaccinated in time $T$, which is fixed and is determined, for example, by the periodically supplied vaccine. In contrast, in the probabilistic scenario $NT\Xi_{\rm pr}=N\int\nolimits_0^T dt\,\xi_{\rm pr}(t)$ gives the average number of vaccinated individuals per time $T$. The actual number is random. For large $NT\Xi_{\rm pr}$, the distribution of the number of vaccinated is close to Gaussian, with variance $NT\Xi_{\rm pr}$.

\subsection{Resonant optimal vaccination}

Expressions (\ref{eq:define_Q}), (\ref{eq:action_w_LS}), and (\ref{eq:action_minimized_explicit}) allow one to investigate the effect of optimal deterministic vaccination on the disease extinction rate, to study how this effect depends on the interrelation between the parameters of the system and vaccination, and to find the change of the disease extinction rate for specific models. The analysis is similar to that for probabilistic vaccination \cite{Khasin2010a}, and we will not reproduce it here. Generally, the effect increases with the increasing period $T$, as seen from Eq.~(\ref{eq:inequal_se}). However, there may be important features that require special attention.

As an illustration we show the results for deterministic vaccination in one of the broadly used epidemic models, the susceptible-vaccinated-infected-recovered (SVIR) model \cite{Andersson2000,Bartlett1960}. Here, in the absence of vaccination, susceptible individuals, with population $X_1=S$, are brought in at rate $\mu N$ (birth), each population decreases at rate $\mu X_i$, $i=1,\ldots,4$ (dying), the infection rate is $\beta X_1X_2/N$, where $X_2=I$ is the number of infected individuals, and infected can recover at rate $\gamma X_2$. The vaccination that we discuss corresponds to the decrease of the number of susceptible individuals at speed $N\xi(t)$.

We assume that both the recovered ($R$) and vaccinated ($V$) individuals keep the immunity, they do not become susceptible again. These groups of population are ``sinks", there is no influx to other groups and the transition rates are independent of $R$ and $V$. The dynamics is then determined by two variables, $X_1=S$ and $X_2=I$. In the mean-field description, for $\beta > \gamma+\mu$ the model possesses a single endemic state $\xb_A$; for $\mu < 4 \,(\beta-\gamma - \mu) (\gamma + \mu)^2\beta^{-2}$ this state is a focus on the plane $(x_1,x_2)$.

\begin{figure}[ht]
\includegraphics[width=3.4in]{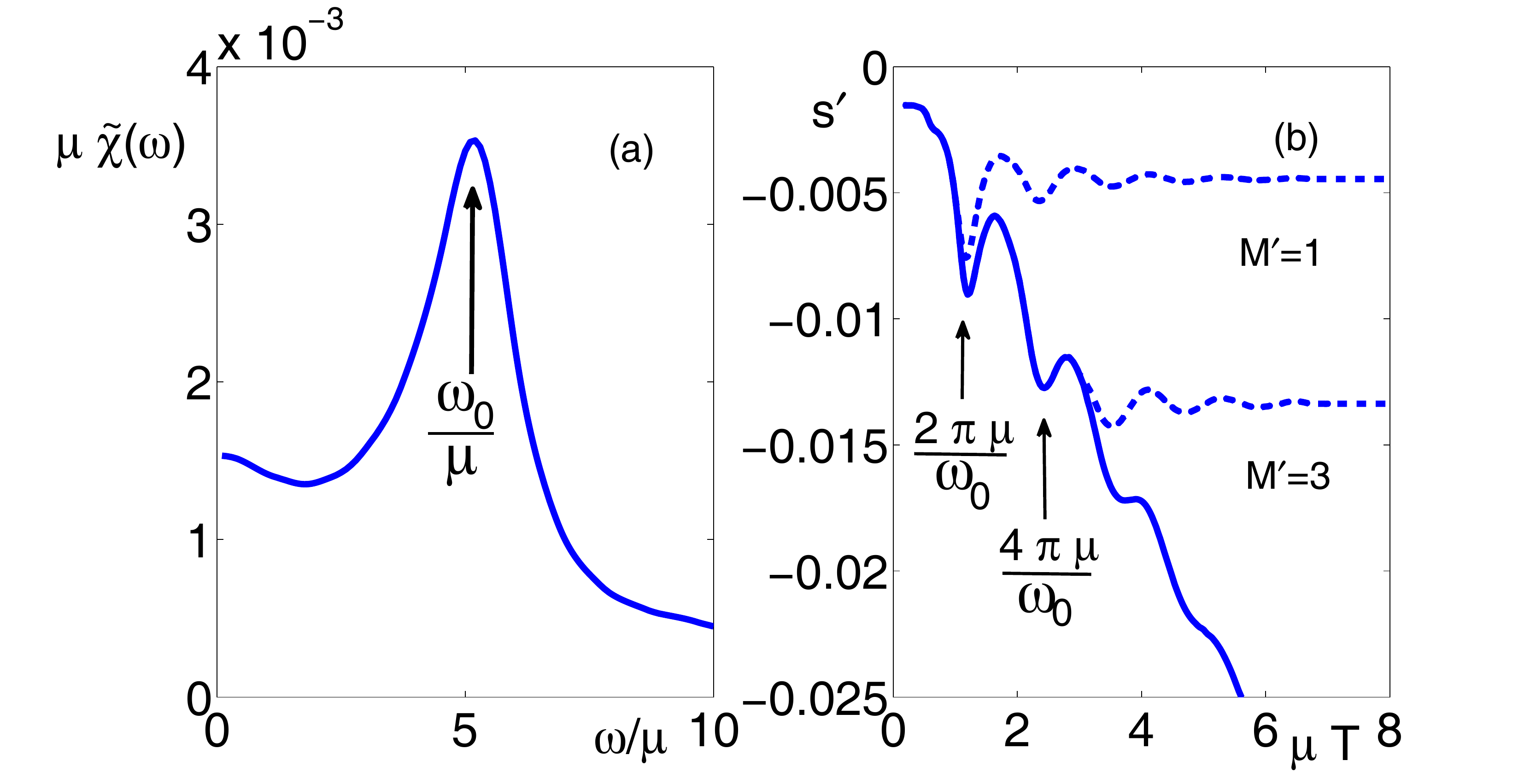}
\caption{(a) The Fourier transform of the logarithmic susceptibility with respect to deterministic vaccination for the SVIR model, Eq.~(\ref{eq:chi_Fourier}). The parameters are $\beta/\mu=80$ and $\gamma/\mu=50$. The susceptibility spectrum displays a sharp peak at the characteristic frequency $\omega_0$ of decaying oscillations near the mean-field endemic state. (b) The change of the scaled extinction barrier $s^{\prime}=\mu \se\1/\Xi$ with vaccination period $T$. The solid line shows $s^{\prime}$ where there is no limit on vaccine accumulation. The dashed lines refer to the case of a limited amount of accumulated vaccine $M$, where the actual mean speed of vaccination is $\Xi_a=\min(\Xi,M/T)$. The scaled accumulation limit is $M'=\mu M/\Xi$. The locations of resonances of $s^{\prime}$ are independent of $M$.}
\label{fig:resonances1}
\end{figure}

In the mean-field approximation, the populations of susceptibles and infected exhibit decaying oscillations in time  as the system approaches the endemic state. In the same parameter range, the populations oscillate also on the optimal disease extinction path \cite{Kamenev2008}. Because of the oscillations, the Fourier-transform of the logarithmic susceptibility
\begin{equation}
\label{eq:chi_Fourier}
\tilde\chi(\omega)=\int\nolimits_{-\infty}^{\infty}dt\,\chi(t)\exp(i\omega t)
\end{equation}
displays a resonant peak shown in Fig.~\ref{fig:resonances1}~(a) [$\tilde\chi(\omega)$ is obtained using the optimal trajectory $\left(\xbo\0(t),\pbo\0(t)\right)$ found in Ref.~\onlinecite{Khasin2010a}]. The peak of $\tilde\chi(\omega)$ is centered at the frequency $\omega_0$ of decaying (in the mean-field approximation) oscillations near $\xb_A$.

Because of the spectral peak in $\tilde\chi(\omega)$, the vaccination-induced term in the extinction exponent ${\cal Q}\1=N\se\1$ depends on the vaccination period $T$ nonmonotonically. This behavior is illustrated in Fig.~\ref{fig:resonances1}~(b). Where $T$ is close to a multiple of the oscillation period $2\pi/\omega_0$, $-\se\1$ displays a peak. Respectively, the disease extinction rate is exponentially enhanced in this case.

Shown in Fig.~\ref{fig:resonances1}~(b) are also the results for $\se\1$ in the case where the total amount of accumulated vaccine is limited, which means $\Xi T\leq M$, where $M$ characterizes the accumulation limit. Such constraint is typical for live vaccine, as it may be dangerous to store too much vaccine in this case. For a given period $T$, this is effectively a constraint on the average vaccination speed $\Xi$, as it makes no sense to increase it beyond $M/T$. If we set $\Xi=M/T$ for large $T$, then it is seen from Eq.~(\ref{eq:inequal_se}) that the effect of vaccination saturates with increasing period,
\[\se\1 \to M\min_t\chi(t), \qquad T\to \infty.\]
However, the maximum of $|\se\1|$ is reached not for $T\to \infty$, as seen from the dashed lines in Fig.~\ref{fig:resonances1}~(b), but for $T$ close to an appropriate multiple of $2\pi/\omega_0$.

If the maximum of $|\se\1|$ lies where the solid and dashed lines are separated, as for $M'=3$, the saturation limit has already been reached for the corresponding $T$, and the actual vaccination speed $\Xi_a=\min(\Xi,M/T)$ is equal to $M/T$. On the other hand, for $M'=1$ the maximum of $|\se\1|$ lies practically on the solid line, and thus it is reached for $\Xi_a \approx \Xi$. In practice, for a given maximum accumulation level $M$ and a given range of available average speeds $\Xi$ and vaccination periods $T$, one should adjust $\Xi$ and $T$ so as to take advantage of the resonance while keeping $\Xi_a$ equal to $\Xi$.

\section{Conclusions}

We have developed a theory of control of large rare fluctuations in reaction and population systems by deterministically imposed transitions. These transitions occur at well-defined instants of time and cause well-defined changes in the numbers of individuals or reacting molecules. The control is comparatively weak, so that the mean-field dynamics of the system is perturbed very weakly. Nevertheless the probabilities of rare events may be changed significantly.

The theory applies to mesoscopic systems, where the characteristic number of molecules or individuals $N$ is large, but not exceedingly large, so that the rates of rare events, which depend on $N$ exponentially, are not exceedingly small. The analysis takes into account the discreteness of the numbers of individuals or molecules and the fact that the tail of the probability distribution is not smooth, the distribution changes by a factor where the number of individuals/molecules changes by one. Our approach is based on the eikonal approximation. It allowed us to reduce the problem of the distribution tail to a Hamilton-Jacobi equation for the action of an auxiliary conservative dynamical system and express the Hamiltonian in terms of the characteristic speed of the deterministic transitions.

Of primary interest for this paper was control of such rare events as switching between coexisting stable states or population extinction. We found that even a comparatively weak control field can lead to an exponential increase of the rates of these events. The exponent is proportional to $N$ and linearly depends on the speed of the imposed transitions. It is determined by the motion of the system in the most probable fluctuation leading to the event in the absence of the control.

The considered control should be contrasted with the ETR control, where all elementary transitions happen at random, but their rates can be controlled. Unexpectedly, we found that the ETR control can be more efficient than the control by deterministically imposed transitions, provided the average transition speed and the composition change in a transition are the same. This is a consequence of the double optimization in the ETR control, where the rate of the rare event is determined by both the optimal fluctuation leading to this event in the absence of control and the optimal realization of the control.

We have considered the problem of optimal control of rare events where a number of deterministically imposed transitions per period is limited. The specific example is disease extinction by vaccination in the situation of a limited average speed of vaccine supply. The optimal shape of the vaccine pulses is a train of $\delta$-pulses. It is independent of the model of the system. The results are illustrated using the well-known susceptible-vaccinated-infected-recovered model, and the possibility of resonant exponential enhancement of the effect of vaccination is demonstrated for the deterministic vaccination.

We note in conclusion that the theory of rare events in mesoscopic population and reaction systems developed in this paper fills the gap between the previously studied deterministic control of rare events in dynamical systems, which are continuous, and the ETR control of reaction systems, which are inherently discrete.

\subsection*{Acknowledgments}
We acknowledge useful discussions with I. B. Schwartz. The work was supported in part by the ARO-57415-NS-II and by NSF Grant PHY-0555346.


%

\end{document}